\PassOptionsToPackage{table,xcdraw}{xcolor}
\documentclass[cameraready]{Interspeech}
\usepackage{multirow}
\usepackage{amsmath}
\usepackage{graphicx}
\usepackage{subcaption}
\usepackage{booktabs} 
\usepackage{arydshln}

\title{Resonate: Reinforcing Text-to-Audio Generation via Online Feedback from Large Audio Language Models}

\author[affiliation={1,2}]{Xiquan}{Li}
\author[affiliation={1}]{Junxi}{Liu}
\author[affiliation={1}]{Wenxi}{Chen}
\author[affiliation={1}]{Haina}{Zhu}
\author[affiliation={1}]{Ziyang}{Ma}
\author[affiliation={1}]{Xie}{Chen}

\address{
    $^1$ X-LANCE Lab, Shanghai Jiao Tong University, China \\
    $^2$ SJTU Paris Elite Institute of Technology, Shanghai Jiao Tong University, China
}

\email{\{mtxiaoxi55, chenxie95\}@sjtu.edu.cn}


\usepackage{comment}

\begin{document}

\maketitle

% the abstract here must exactly match the abstract entered into the paper submission system
\begin{abstract}
Reinforcement Learning (RL) has become an effective paradigm for enhancing Large Language Models (LLMs) and visual generative models. However, its application in text-to-audio (TTA) generation remains largely under-explored. Prior work typically employs offline methods like Direct Preference Optimization (DPO) and leverages Contrastive Language-Audio Pretraining (CLAP) models as reward functions. In this study, we investigate the integration of online Group Relative Policy Optimization (GRPO) into TTA generation. We adapt the algorithm for Flow Matching-based audio models and demonstrate that online RL significantly outperforms its offline counterparts. Furthermore, we incorporate rewards derived from Large Audio Language Models (LALMs), which can provide fine-grained scoring signals that are better aligned with human perception. With only 470M parameters, our final model, \textbf{Resonate}, establishes a new SOTA on TTA-Bench in terms of both audio quality and semantic alignment.\footnote{Demo: \url{https://resonatedemo.github.io}. Code: \url{https://github.com/xiquan-li/Resonate}.}
    % 1000 characters. ASCII characters only. No citations.
\end{abstract}

\section{Introduction}
Text-to-audio (TTA) \cite{liu2023audioldm, huang2023make, ghosal2023text} generation aims to synthesize high-fidelity and diverse acoustic signals guided by natural language descriptions.
With the advancements in architectural design \cite{evans2025stable, guan2024lafma, hung2024tangoflux}, data curation \cite{bai2025audiosetcaps, kong2024improving}, and training objectives \cite{majumder2024tango2, hung2024tangoflux, li2025meanaudio}, TTA models have recently achieved significant improvements. 
These advances have enabled the seamless generation of complex soundscapes and effects, facilitating automated content creation across filmmaking, gaming, and virtual reality.

Reinforcement learning (RL) has established itself as an effective paradigm to align deep neural networks with human perception.
In particular, it has shown promising results in various domains, including Large Language Models \cite{shao2024deepseekmath}, Image Generation \cite{liu2025flow, li2025mixgrpo}, Text-to-speech Synthesis \cite{sun2025f5r}, and Speech enhancement \cite{wang2026flowse}. 
In the realm of Text-to-Audio (TTA) generation, researchers have also actively explored RL-based techniques to enhance the model's controllability and synthesis quality.
For instance, BATON \cite{liao2024baton} pioneered this direction by training a custom reward model on a human-annotated audio dataset and performing weighted supervised fine-tuning on top of the data labeled by the reward model. Tango 2 \cite{majumder2024tango2} further employs the CLAP model \cite{elizalde2023clap, wu2023large} to curate a preference dataset, and employs direct preference optimization (DPO) \cite{rafailov2023direct} to enhance the model's capabilities. T2A-Feedback \cite{wang2025t2a} further advances the preference data curation pipeline by incorporating source-separation models, audio-grounding models, and quality-assessment models to create better preference data pairs. 
More recently, TangoFlux \cite{hung2024tangoflux} presents a novel CLAP-Ranked Preference Optimization (CRPO) framework, where they utilize an off-the-shelf CLAP model as the reward model and perform iterative DPO to enhance the model's capability.

Despite these initial successes, existing RL-based TTA methods still face two inherent limitations. First, they predominantly adopt offline RL paradigms such as Direct Preference Optimization (DPO). Nonetheless, the decoupled process of preference data generation and model training in offline RL can lead to distribution shift and limit the exploration of the policy model. 
Secondly, existing frameworks predominantly rely on CLAP as their reward model. However, CLAP models suffer from a 'bag-of-words' effect \cite{ghosh2023compa, yuan2024t}, limiting their temporal and compositional reasoning and yielding coarse-grained rewards that poorly align with human preferences \cite{kuan2026aqascore}.

In this paper, we propose integrating online RL methods Group Relative Policy Optimization (GRPO) \cite{shao2024deepseekmath, liu2025flow} into TTA training. By adapting the algorithm for Flow Matching-based TTA models, we demonstrate significant improvements in both semantic consistency and audio generation quality.
Complementing this, we employ Large Audio Language Models (LALMs) \cite{bai2025qwen25omni, xu2025qwen3} as reward models, utilizing their sophisticated reasoning and perception abilities to generate fine-grained feedback that aligns better with human judgment. 
% We demonstrate that the reward signals from LALM can effectively improve model's capability. 
Finally, we present \textbf{Resonate}, a novel text-to-audio generator that achieves state-of-the-art (SOTA) performance on the TTA-Bench dataset, validating the efficacy of our designs.
Our main contributions are summarized as follows:
\begin{itemize}
    \item \textbf{Online RL for TTA:} We present the first successful integration of online RL (GRPO) into the text-to-audio generation pipeline, enabling superior optimization efficiency.
    \item \textbf{LALM as Reward Model:} We leverage modern LALMs to provide fine-grained reward signals that align better with human preferences. 
    \item \textbf{SOTA Performance:} We introduce \textbf{Resonate}, a novel high-fidelity generator that achieves state-of-the-art performance on TTA-Bench in both audio quality and semantic alignment.
\end{itemize}
We will open-source the code and model weights to promote future research in the community.

\section{Methods}

\begin{figure*}
    \centering
    \includegraphics[width=\linewidth]{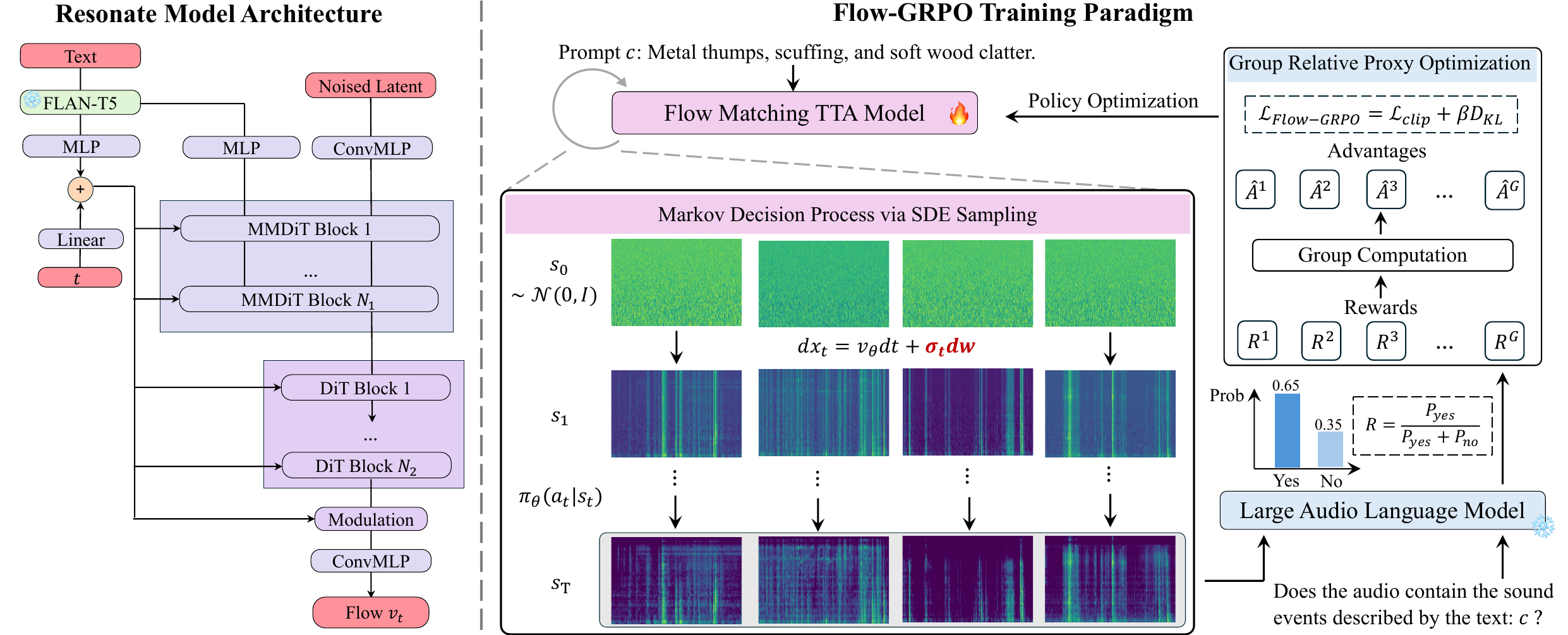}
    \caption{Left: the model architecture of Resonate. Right: The employed Flow-GRPO training paradigm. }
    \label{fig:resonate_model}
    \vspace{-0.5cm}
\end{figure*}

\subsection{Model Architecture}
\label{sec:model_architecture}
Following MeanAudio \cite{li2025meanaudio}, Resonate employs a Flux-style flow Transformer \cite{flux2024} to model the generative process in the latent space. 
Given an audio waveform $a$ and its corresponding textual description $c$, we first convert the waveform into a mel-spectrogram and encode it into VAE latents $x$. 
In parallel, the textual description is encoded into conditioning embeddings using the FLAN-T5 text encoder \cite{chung2024scaling}.
The noised audio latent $x_t$ and the encoded text embeddings are fed into the backbone Transformer, which consists of $N_1$ multi-modal (MMDiT) blocks \cite{esser2024scaling} followed by $N_2$ single-modal (DiT) blocks \cite{peebles2023scalable}. 

Before introducing reinforcement learning, we first pre-train the model using the Conditional Flow Matching (CFM) objective \cite{liu2022flow}, where the network learns to predict the instantaneous velocity field of the probability flow. The objective is defined as:
$$
\mathcal{L}_{\text{CFM}} 
= \mathbb{E}_{t,x,\epsilon}
\lVert v_\theta(x_t, t, c) - v_t  \rVert_2^2 ,
$$
where $x_t = (1 - t)x + t\epsilon$ denotes the noised audio latent, and the target velocity is given by
$
v_t = \frac{d x_t}{dt} = \epsilon - x
$.

\begin{table*}[htbp]
\centering
\caption{Evaluation results on the Accuracy subset of TTA-Bench}
\vspace{-0.2cm}
\label{tab:tta_bench_results}
\resizebox{\textwidth}{!}{
\begin{tabular}{lccccccccccc}
\toprule
\multirow{2.5}{*}{\textbf{Model}} & \multirow{2.5}{*}{\textbf{Params}} & \multirow{2.5}{*}{\textbf{NFE}} & \multicolumn{6}{c}{\textbf{Objective Metrics}} & \multicolumn{2}{c}{\textbf{Subjective Metrics}} \\
\cmidrule(lr){4-9} \cmidrule(lr){10-11} 
& & & \textbf{CE} $\uparrow$ & \textbf{CU} $\uparrow$ & \textbf{PC} $\uparrow$ & \textbf{PQ} $\uparrow$ & \textbf{CLAP} $\uparrow$ & \textbf{AQAScore} $\uparrow$ & \textbf{OVL} & \textbf{REL} \\
\midrule
AudioLDM-L-Full \cite{liu2023audioldm}  & 739M  & 200 & 3.275 & 5.137 & 3.219 & 5.854 & 0.441 & 0.574 & - & - \\
AudioLDM-2-Large \cite{liu2024audioldm2}  & 718M  & 200 & 3.441 & \textbf{5.440} & 2.957 & \underline{5.984} & 0.415 & 0.589 & - & - \\
Tango-Full  \cite{ghosal2023text}   & 866M  & 200 & 3.264 & 5.151 & 3.360 & 5.954 & 0.440 & 0.599 & - & - \\
Tango-2-Full  \cite{majumder2024tango2}    & 866M  & 200 & \underline{3.468} & 5.196 & \textbf{3.800} & 5.895 & 0.467 & 0.702 & - & - \\
MeanAudio-L-Full \cite{li2025meanaudio}& 480M  & 25  & 3.304 & 5.151 & 2.995 & 5.743 & 0.469 & \underline{0.729} & - & - \\
EzAudio-XL   \cite{hai2024ezaudio}    & 857M  & 200 & 3.388 & 5.055 & 3.663 & 5.719 & 0.446 & 0.709 & \underline{3.54}\scriptsize{$\pm$0.79} & 3.53\scriptsize{$\pm$0.89}\\
GenAU-L-Full  \cite{haji2024taming}   & 1253M & 200 & 3.359 & 5.096 & 3.584 & 5.802 & 0.468 & 0.646 & 3.52\scriptsize{$\pm0.83$} & 3.37\scriptsize{$\pm0.88$} \\
TangoFlux   \cite{hung2024tangoflux}     & 516M  & 50  & \textbf{3.539} & 5.074 & \underline{3.673} & 5.782 & \underline{0.472} & 0.677 & 3.51\scriptsize{$\pm0.86$} & \underline{3.66}\scriptsize{$\pm0.83$} \\
\midrule
Resonate-PT  (Ours)  & 470M  & 25  & 3.408 & 5.238 & 3.002 & 5.923 & 0.458 & 0.651 & 3.44\scriptsize{$\pm0.93$} & 3.37\scriptsize{$\pm0.89$}\\
Resonate-GRPO (Ours) & 470M  & 25  & 3.451 & \underline{5.328} & 3.232 & \textbf{6.064} & \textbf{0.476} & \textbf{0.737} & \textbf{3.86\scriptsize{$\pm0.81$}} & \textbf{3.83\scriptsize{$\pm0.83$}} \\
\bottomrule
\end{tabular}
}
\vspace{-0.3cm}
\end{table*}

\subsection{Flow-GRPO}
Following \cite{black2023training, liu2025flow}, we formalize the iterative denoising process of flow matching models as a Markov Decision Process (MDP), denoted by the tuple $(\mathcal{S}, \mathcal{A}, \rho_0, P, R)$. At timestep $t$, we define the state as $s_t \triangleq (c, t, x_t)$ and the action as the predicted next sample $a_t \triangleq x_{t-1}$. Accordingly, the policy is given by $\pi(a_t \mid s_t) \triangleq p_\theta(x_{t-1} \mid x_t, c)$. The transition dynamics are deterministic, satisfying $P(s_{t+1} \mid s_t, a_t) \triangleq (\delta_c, \delta_{t-1}, \delta_{x_{t-1}})$. The initial state distribution is $\rho_0(s_0) \triangleq (p(c), \delta_T, \mathcal{N}(0, I))$, where $\delta_y$ represents the Dirac delta distribution centered at $y$. Finally, the reward is assigned only at the terminal step: $R(s_t, a_t) \triangleq r(x_0, c)$ if $t=0$, and $0$ otherwise.

Based on the MDP formulation, Flow-GRPO \cite{liu2025flow} adapts the Group Relative Policy Optimization (GRPO) \cite{shao2024deepseekmath} algorithm to flow matching models. For each prompt $c$, the policy $\pi_{\theta_{\text{old}}}$ samples a group of $G$ trajectories, denoted as $\{( x_T^i, x_{T-1}^i, \dots, x_0^i )\}_{i=1}^G$, where $x_0^i$ represents the final generated audio for the $i$-th sample. The advantage $\hat{A}^i$ for each sample in the group is estimated by normalizing the final rewards:
$$
\hat{A}^i = \frac{R(x_0^i, c) - \text{mean}(\{R(x_0^j, c)\}_{j=1}^G)}{\text{std}(\{R(x_0^j, c)\}_{j=1}^G)}.
$$
The policy is then optimized by maximizing the following surrogate objective, which averages the clipped policy gradients over the group:
\begin{equation}
% \label{eq:loss_flowgrpo}
    \mathcal{J}_{\text{Flow-GRPO}}(\theta) = \mathbb{E}_{c \sim \mathcal{C}, \{x^i\}_{i=1}^G \sim \pi_{\theta_{\text{old}}}} \left[ \frac{1}{G} \sum_{i=1}^{G} \frac{1}{T} \sum_{t=0}^{T-1} \mathcal{L}_t^i(\theta) \right],
\end{equation}
where the per-step loss term $\mathcal{L}_t^i(\theta)$ is defined as:
\begin{equation*}
\begin{aligned}
    \mathcal{L}_t^i(\theta) &= \min \left( r_t^i(\theta) \hat{A}^i, \text{clip}\left(r_t^i(\theta), 1-\varepsilon, 1+\varepsilon\right) \hat{A}^i \right) \nonumber \\
    &\quad - \beta D_{\text{KL}}\left(\pi_\theta \parallel \pi_{\text{ref}}\right).
\end{aligned}
\end{equation*}
Here, $r_t^i(\theta)$ represents the probability ratio between the current and old policies:
$
    r_t^i(\theta) 
    % = \frac{\pi_\theta(x_{t-1}^i \mid s_t^i)}{\pi_{\theta_{\text{old}}}(x_{t-1}^i \mid s_t^i)} 
    = \frac{p_\theta(x_{t-1}^i \mid x_t^i, c)}{p_{\theta_{\text{old}}}(x_{t-1}^i \mid x_t^i, c)}.
$

Since GRPO requires sampling diverse audios through stochastic exploration, the standard flow-matching denoising process is inherently unsuitable as it relies on a deterministic ODE sampler:
\begin{equation*}
    dx_t = v_\theta(x_t, t, c) dt
\end{equation*}
To introduce the necessary stochasticity, Flow-GRPO converts this deterministic ODE into an equivalent SDE sampler that preserves the same marginal probability densities at all time steps. The resulting reverse-time SDE is given by:
\begin{equation*}
    dx_t = \left( v_t(x_t) - \frac{\sigma_t^2}{2} \nabla \log p_t(x_t) \right) dt + \sigma_t dw.
\end{equation*}
Here, $dw$ denotes Wiener process increments and $\sigma_t=a\sqrt{\frac{t}{1-t}}$ controls the level of stochasticity during generation. 
The above expression can be further transformed into: 
\begin{equation*}
    dx_t = \left[ v_t(x_t) + \frac{\sigma_t^2}{2(1-t)}(x_t + t v_t(x_t)) \right] dt + \sigma_t dw.
\end{equation*}
This continuous SDE leads to the final discretized update rule used during the transition steps:
\begin{align*}
    x_{t+\Delta t} &= x_{t,\text{mean}} + \sigma_t \sqrt{|\Delta t|} \epsilon, \\
    x_{t,\text{mean}} &= x_t + \left[ v_\theta(x_t, t) + \frac{\sigma_t^2}{2t}(x_t + (1-t)v_\theta(x_t, t)) \right] \Delta t.
\end{align*}
Under this stochastic updating scheme, the transition likelihood $p_\theta(x_{t-1}^i \mid x_t^i, c)$ utilized in the GRPO objective is explicitly defined as the Gaussian density of $x_{t,\text{mean}}$ with mean $x_{t,\text{mean}}$ and standard deviation $\sigma_t \sqrt{|\Delta t|}$.
By maximizing the surrogate objective $\mathcal{J}_{\text{Flow-GRPO}}(\theta)$, the algorithm effectively updates the policy model by optimizing these transition probabilities.

% {"summary": {"REL": {"LCC": 0.5757637692446425, "SRCC": 0.5868045402696297, "KTAU": 0.4290912512121245}, "OVL": {"LCC": 0.4036501839206115, "SRCC": 0.4064996566921892, "KTAU": 0.2857135228746575}}}

\begin{table}[t]
\centering
\caption{
Correlation comparison on the the PAM dataset
}
\label{tab:correlation}
\vspace{-0.2cm}
\setlength{\tabcolsep}{8pt}
\resizebox{\linewidth}{!}{
\begin{tabular}{lccc}
\toprule
\textbf{Models} & \textbf{LCC$\uparrow$} & \textbf{SRCC$\uparrow$} & \textbf{KTAU$\uparrow$} \\
\midrule
 LAION-CLAP-630K-Best & 0.472 & 0.477 & 0.337 \\
 Qwen2.5-Omni-7B & 0.518 & \textbf{0.589} & \textbf{0.429} \\
 Qwen3-Omni-Instruct & \textbf{0.575} & 0.586 & \textbf{0.429} \\
\bottomrule
\end{tabular}
}
\vspace{-0.5cm}
\end{table}

% \begin{table}[t]
% \centering
% \caption{
% Evaluation results on AudioCaps test set
% }
% \label{tab:results_audiocaps}
% \vspace{-0.2cm}
% \setlength{\tabcolsep}{8pt}
% \resizebox{0.9\linewidth}{!}{
% \begin{tabular}{lccc}
% \toprule
% \textbf{Models} & \textbf{FAD$\downarrow$} & \textbf{KL$\downarrow$} & \textbf{CLAP$\uparrow$} \\
% \midrule
% GenAU-L-Full \cite{haji2024taming} &  \textbf{2.07} & 1.36 & 0.300 \\
% TangoFlux \cite{hung2024tangoflux}  &  2.41 & \textbf{1.27} & 0.318 \\
% \hdashline
% Resonate - PT (Ours)  & 3.19 & 1.35 & 0.395 \\ 
% Resonate - GRPO (Ours)  & 2.53 & \textbf{1.27}  & \textbf{0.416}
% \\
% \bottomrule
% \end{tabular}
% }
% \vspace{-0.5cm}
% \end{table}

\subsection{Reward from LALM}
\label{sec:reward_lalm}
Given the robust auditory understanding and reasoning capabilities demonstrated by recent Large Audio-Language Models (LALMs) \cite{chu2024qwen2, ghosh2025audio}, we employ them as the reward models for our GRPO training.
Following AQAScore \cite{kuan2026aqascore}, we frame the reward evaluation as an Audio Question Answering (AQA) task. Specifically, for a given generated audio $a$ and its corresponding text prompt $c$, we construct a query $q(c)$ formatted as: \texttt{``Does the audio contain the sound events described by the text: \{c\}? Please answer yes or no.''}

Let $p_\psi(y \mid a, q(c))$ denote the conditional probability of the response generated by an LALM parameterized by $\psi$, where the target answer $y \in \{\text{``Yes''}, \text{``No''}\}$. We define the reward $R(a, c)$ as the normalized probability of the affirmative response. By denoting the log-likelihood for each response as $s_y = \log p_\psi(y \mid a, q(c))$, the reward is formulated via a softmax function:
\begin{equation*}
    R(a, c) = \frac{\exp(s_{\text{yes}})}{\exp(s_{\text{yes}}) + \exp(s_{\text{no}})}.
\end{equation*}
As demonstrated in \cite{kuan2026aqascore}, this AQA-based evaluation (denoted as AQAScore) can capture fine-grained audio details and correlates strongly with human subjective judgments. Moreover, it is computationally efficient, as the model is constrained to output only a single token.
Table \ref{tab:correlation} reports representative correlations between model-assigned scores and human ratings on the PAM dataset \cite{deshmukh2024pam}. The results for Qwen2.5-Omni \cite{bai2025qwen25omni} and LAION-CLAP \cite{wu2023large} are taken from \cite{kuan2026aqascore}, while the performance of Qwen3-Omni-Instruct \cite{xu2025qwen3} is evaluated by ourselves. 

\section{Experiments}

\subsection{Datasets}
We curated a large-scale audio-text corpus to pre-train the model with the CFM objective before applying Flow-GRPO. This corpus comprises approximately 3.7 million audio-text pairs, totaling 10,000 hours of audio data, sourced from AudioCaps \cite{kim2019audiocaps}, AudioSet \cite{gemmeke2017audio, bai2025audiosetcaps}, Clotho \cite{drossos2020clotho}, VGGSound \cite{chen2020vggsound}, WavCaps \cite{mei2024wavcaps}, and AudioStock\footnote{\url{https://audiostock.net/}}. Following the pre-training phase, we optimize the model using Flow-GRPO on the prompts set of AudioCaps training split.

% For evaluation, we use TTA-Bench \cite{wang2025tta} and AudioCaps. TTA-Bench is a recent benchmark designed for comprehensive evaluation of audio generative models. We adopt its Accuracy subset, which contains 1,500 diverse prompts with multiple sound events and complex temporal relationships, providing a challenging test of instruction-following ability. 
% We also evaluate models on the AudioCaps test set, which includes 957 audio samples.

For evaluation, we employ TTA-Bench \cite{wang2025tta}, which is a recently developed benchmark specifically designed for the comprehensive assessment of audio generative models. We adopt the Accuracy subset of TTA-Bench, which provides 1,500 diverse prompts spanning various real-world scenarios. These prompts feature multiple sound events and complex temporal logic (parallel, sequential, and complex interactions), offering a robust test for the model's instruction-following capabilities across intricate acoustic scenes.

\subsection{Training and Evaluation Details}
Resonate comprises $N_1=16$ MMDiT blocks and $N_2=36$ DiT blocks, the hidden dimension is set to 448 and the model contains 470M parameters in total. 
We pre-train the model with a batch size of 256 for 500k steps. 
During post-training, we apply Flow-GRPO for 1,000 steps, using a group size $G=24$, a noise level $a=0.7$, and a KL coefficient $\beta=0.04$.
We use Qwen2.5-Omni \cite{bai2025qwen25omni} as the reward model, and use the methods described in Section \ref{sec:reward_lalm} to calculate reward. 

For objective evaluation, we employ AudioBox-Aesthetics \cite{tjandra2025meta} to assess the quality of generated audio. This model provides four distinct metrics capturing various acoustic dimensions: Content Enjoyment (CE), Content Usefulness (CU), Production Complexity (PC), and Production Quality (PQ). To evaluate semantic alignment, we report the CLAP score computed with Microsoft CLAP \cite{elizalde2023clap}, together with AQAScore (detailed in Section \ref{sec:reward_lalm}) computed using Qwen3-Omni-Instruct. 
Note that we deliberately use different models for reward optimization (Qwen2.5-Omni) and evaluation (Qwen3-Omni-Instruct) to reduce the risk of reward hacking.
For subjective evaluation, we sample 10 generated waveforms for assessment by a panel of 10 audio experts. The experts rate each sample on a scale of 1 to 5 based on two criteria: Overall Quality (OVL) and Relevance (REL) to the input text.

\subsection{Main Results}
Table \ref{tab:tta_bench_results} presents the quantitative results on the TTA-Bench, where Resonate-PT denotes our pre-trained Flow Matching model and Resonate-GRPO represents the reinforced variant. 
Resonate-GRPO consistently achieves state-of-the-art performance, with particularly notable gains in semantic alignment. 
Specifically, our model attains a dominant AQAScore of 0.737, surpassing both the previous state-of-the-art, MeanAudio (0.729), and the robust Flow Matching baseline, TangoFlux (0.677), while simultaneously securing the highest CLAP score of 0.476.
Regarding audio fidelity, Resonate-GRPO excels in Production Quality (PQ), a metric evaluating perceptual fidelity, with a top score of 6.064, while maintaining highly competitive Content Usefulness (CU) at 5.328.
Although AudioLDM-2-Large achieves a marginally higher CU, it significantly underperforms in semantic alignment (AQAScore: 0.589), underscoring our model’s superior equilibrium between content validity and text-audio adherence. 
Crucially, the comparison against the pre-trained baseline (Resonate-PT) confirms that Flow-GRPO yields comprehensive improvements across all dimensions: it significantly boosts semantic alignment (AQAScore: $0.651 \rightarrow 0.737$) while concurrently enhancing audio fidelity (PQ: $5.923 \rightarrow 6.064$), demonstrating that our RL strategy effectively refines both generation quality and semantic accuracy. 
The subjective evaluation further confirms this superiority, as Resonate-GRPO achieves the highest Overall Quality (OVL: 3.86) and Relevance (REL: 3.83) scores among all compared models.
Notably, these superior results are achieved with a remarkably compact architecture (470M parameters) and highly efficient inference (25 NFE).

\subsection{Ablation Studies}

\begin{table}[t]
\centering
\caption{
Ablation study on different training paradigms. 
}
\label{tab:ablation_training_paradigm}
\vspace{-0.2cm}
\setlength{\tabcolsep}{8pt}
\resizebox{\linewidth}{!}{
\begin{tabular}{lcccc}
\toprule
\textbf{Methods} & \textbf{CU$\uparrow$} & \textbf{PQ$\uparrow$} & \textbf{CLAP$\uparrow$} & \textbf{AQAScore$\uparrow$} \\
\midrule
Resonate-PT & 5.238	& 5.923& 	0.458&	0.651 \\
% \quad w. DPO ($\beta=100$) & 5.239	& 5.921	& 0.464 & 0.663 \\
\quad w. DPO ($\beta=100$) & 5.237	& 5.924	& 0.462 & 0.676 \\

\quad w. SFT & 5.114	& 5.764	& 0.461 &	0.675   \\
\quad w. SFT + GRPO  & 5.184 &	5.883	& 0.472	& 0.728 \\
\rowcolor{blue!10}
\quad w. GRPO & \textbf{5.328}	& \textbf{6.064}	& \textbf{0.476} &	\textbf{0.737} \\
\bottomrule
\end{tabular}
}
% \vspace{-0.3cm}
\end{table}

\begin{table}[t]
\centering
\caption{
Ablation study on different reward models. 
}
\label{tab:ablation_reward_models}
\vspace{-0.2cm}
\setlength{\tabcolsep}{8pt}
\resizebox{\linewidth}{!}{
\begin{tabular}{lcccc}
\toprule
\textbf{Methods} & \textbf{CU$\uparrow$} & \textbf{PQ$\uparrow$} & \textbf{CLAP$\uparrow$} & \textbf{AQAScore$\uparrow$} \\
\midrule
Resonate-PT & 5.238	& 5.923 & 0.458 & 0.651 \\
\quad w. CLAPScore & 5.276 &	6.063	& \textbf{0.479} &	0.711  \\
\rowcolor{blue!10}
\quad w. AQAScore  & \textbf{5.328}	& \textbf{6.064}	&  0.476 &	\textbf{0.737}  \\
\bottomrule
\end{tabular}
}
\vspace{-0.4cm}
\end{table}

We conduct a comprehensive ablation study to validate the effectiveness of our design components. 

\noindent \textbf{Training Strategies.}
We first investigate the effectiveness of different training strategies. To this end, we compare our approach with the offline RL method Direct Preference Optimization (DPO) \cite{wallace2024diffusion}, Supervised Fine-Tuning (SFT), and SFT followed by GRPO. For SFT, we fine-tune the model on the AudioCaps training set for 50k steps. AudioCaps is selected as it represents the largest audio dataset with high-quality human-annotated captions.
As illustrated in Table \ref{tab:ablation_training_paradigm}, the offline RL baseline (DPO) yields only marginal gains over the pre-trained model (Resonate-PT), with the AQA Score rising slightly (0.651 to 0.676) while audio quality metrics (CU/PQ) remain stagnant (5.237/5.924).
In stark contrast, our proposed GRPO achieves a substantial boost across all metrics, outperforming DPO by a significant margin (e.g., AQA Score: 0.737 vs. 0.676), showing the advantages of online RL.

Although SFT improves semantic alignment (CLAP to 0.461, AQAScore to 0.675), it significantly degrades audio quality: CU drops from 5.238 (PT) to 5.114, and PQ decreases to 5.764. We attribute this to the “in-the-wild” nature of AudioCaps, which contains recordings with variable and often low fidelity. While high-quality human captions enhance semantic alignment, the model overfits to the dataset’s noisy acoustic characteristics, leading to deteriorated generation quality.
Notably, the multi-stage strategy (SFT followed by GRPO) underperforms directly applying GRPO to the pretrained model. Although SFT+GRPO partially recovers performance relative to SFT alone, it remains inferior to GRPO-only, which achieves the best overall results. 
We hypothesize that SFT restricts policy exploration by overfitting to the noisy AudioCaps distribution, thereby limiting the effectiveness of subsequent RL optimization. 
This highlights the critical role of data quality in TTA training and suggests a need for constructing higher-quality datasets to better support supervised fine-tuning.

\noindent \textbf{Reward Models. }
To validate the effectiveness of the proposed LALM-based reward, we compare it with directly using CLAP as the reward model. As shown in Table \ref{tab:ablation_reward_models}, using AQAScore as the reward leads to better performance across a wide range of metrics.
Specifically, using AQAScore as the reward achieves the highest audio quality (CU: 5.328, PQ: 6.064) and the best human-aligned semantic preference (AQAScore: 0.737). 
Although optimizing CLAP as the reward yields a slightly higher CLAP score (0.479 vs. 0.476) due to direct objective optimization, it fails to deliver comparable overall improvements. 
These results indicate that LALM can provide more informative feedback for audio generation, leading to superior fidelity.

\begin{figure}[t]
    \centering
    \includegraphics[width=\linewidth]{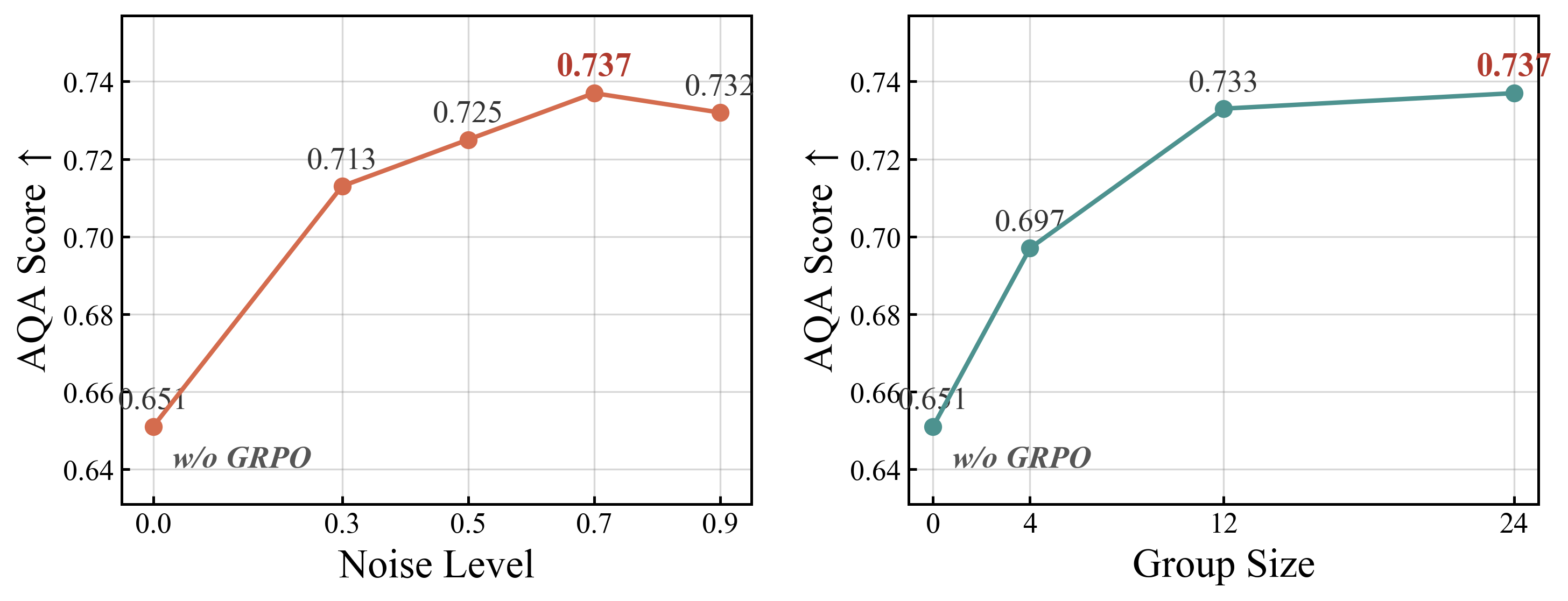}
    \vspace{-0.5cm}
    \caption{Ablation study on Flow-GRPO configurations. }
\label{fig:ablation_config}
\vspace{-0.6cm}
\end{figure}

\noindent \textbf{Flow-GRPO Configuration. }
We further investigate the key hyperparameters in GRPO training, specifically the noise level parameter ($a$ in $\sigma_t$) and the group size $G$.
Regarding the noise level (Figure \ref{fig:ablation_config}a), introducing noise is essential for effective exploration, with performance improving as the noise increases and peaking at 0.7 with an AQA Score of 0.737.
% This suggests that moderate noise promotes exploration and helps find the optimal policy. 
However, excessive noise (e.g., 0.9) degrades performance (0.732) and can easily lead to reward hacking, suggesting that moderate noise helps find the optimal policy.
For the group size (Figure \ref{fig:ablation_config}b), performance increases monotonically from 4 to 24, suggesting that larger groups can provide more reliable advantage estimates and can stabilize optimization. 
We select a group size of 24 as a practical trade-off between performance and efficiency.

\section{Conclusion}
This paper studies the integration of online reinforcement learning into TTA generation. By adapting GRPO to flow-matching-based audio generation models, we demonstrate that online RL yields substantial improvements over conventional offline methods. 
To better align rewards with human preferences, we leverage the perceptual and reasoning capabilities of LALMs, reformulating audio evaluation as a question-answering task.
Building on these components, we introduce Resonate, a novel text-to-audio generator that achieves SOTA performance in both audio quality and semantic alignment on the TTA-Bench.

% \newpage
\section{Generative AI Use Disclosure}
In the preparation of this manuscript, AI-based writing tools were employed exclusively to enhance linguistic quality, including grammatical correction and stylistic refinement. The authors maintain full responsibility for the research integrity; no generative AI was used to create original data, experimental results, or scientific conclusions. All intellectual contributions, including the core methodology and analysis, were independently conceived and executed by the research team.

\bibliographystyle{IEEEtran}
\bibliography{mybib}

\end{document}